\documentclass[a4paper,11pt]{article}

\usepackage{jheppub} 

\usepackage[T1]{fontenc} 

\usepackage{simplewick}

\newcommand{\beq}{\begin{eqnarray}}
\newcommand{\eeq}{\end{eqnarray}}

\usepackage{graphicx}
\usepackage{amssymb}
\usepackage{amsmath}
\usepackage{amsfonts}
\usepackage{bbm}

\usepackage{color}

\title{\boldmath Fractional instanton of the SU($3$) gauge theory in weak coupling regime}

\author[a,b,c,1]{Etsuko Itou,\note{Corresponding author.}}

\affiliation[a]{Department of Physics, and Research and Education Center for Natural Sciences, Keio University, 4-1-1 Hiyoshi, Yokohama, Kanagawa 223-8521, Japan}
\affiliation[b]{Department of Mathematics and Physics, Kochi University, Kochi 780-8520, Japan}
\affiliation[c]{Research Center for Nuclear Physics (RCNP), Osaka University, Osaka 567-0047, Japan}

\emailAdd{itou@yukawa.kyoto-u.ac.jp}

\abstract{Motivated by recent studies on the resurgence structure of quantum field theories, we numerically study the nonperturbative phenomena of the SU($3$) gauge theory in a weak coupling regime. 
We find that topological objects with a fractional charge emerge if the theory is regularized by an infrared (IR) cutoff via the twisted boundary conditions.
Some configurations with nonzero instanton number are generated as a semi-classical configuration in the Monte Carlo simulation even in the weak coupling regime.
Furthermore, some of them consist of multiple fractional-instantons.
We also measure the Polyakov loop to investigate the center symmetry and confinement.
The fractional-instanton corresponds to a solution linking two of degenerate $\mathbb{Z}_3$-broken vacua in the deconfinement phase.}

\begin{document} 
\maketitle
\flushbottom

\section{Introduction}
Instanton is one of the classical solutions of the quantum field theory and labels a vacuum state.
In SU($N_c$) gauge theories in the four-dimensional Euclidean spacetime (${\mathbb{R}^4}$), an instanton has a topological charge (instanton number) given by
\beq
Q = \frac{1}{32 \pi^2} \int d^4 x \mbox{Tr} \epsilon_{\mu \nu \rho \sigma}  F_{\mu \nu} F_{\rho \sigma}, \label{eq:def-Q}
\eeq
where $\epsilon_{\mu \nu \rho \sigma}$ denotes a totally anti-symmetric tensor.
This $Q$ always takes an integer value~\cite{Atiyah:1968mp}.
The topological charge has been measured by lattice {\it ab initio} calculations~\cite{Luscher:1981zq}.
The distribution of $Q$ is broad in the hadronic (confinement) phase in the low temperature, while it is narrow in the quark-gluon-plasma (deconfinement) phase in the high temperature~\cite{Alles:1996nm, Durr:2006ky}. 
According to lattice numerical studies~\cite{Kogut:1982rt, Fukugita:1989yb,Fukugita:1989yw}, it has been shown that the first-order phase transition occurs between these phases in the SU($3$) gauge theory, so that physical observables do not continuously change around the critical temperature.

Most calculations of lattice at zero temperature have been performed on the hypertorus ($\mathbb{T}^4$) with the standard periodic boundary conditions, although the gauge theories on the periodic hypertorus have neither self-dual nor anti-self-dual configuration as a classical solution.
To obtain a stable configuration with nonzero topological charge on the finite lattice, we have to impose the {\it twisted} boundary conditions~\cite{tHooft:1981nnx}.
The reason why the ordinary lattice calculations with the periodic boundary conditions can observe the nonzero $Q$ is that the boundary effect is negligible in the strong coupling regime.

A question arises as to properties of the topological objects in the weak coupling regime of the SU($3$) gauge theory. 
To study the quantities in the perturbative regime on the lattice, for instance, to calculate the running coupling constant, we need to set the renormalization scale to be higher than the Lambda scale ($\Lambda$).
The renormalization scale is inversely proportional to the spatial lattice extent ($L_s$), hence we have to use the lattice extent satisfying $L_s \ll 1/ \Lambda$.
We expect that the choice of the boundary condition in such a small box has an influence on the property of classical solutions.
Therefore, it is necessary to consider which are a proper boundary condition and spacetime structure to investigate the weak coupling regime.

As a proper boundary condition to match the lattice data with the perturbative calculations on $\mathbb{R}^4$, it is necessary to utilize nontrivial boundary conditions ({\it{e.g.}} the Schr\"{o}dinger functional~\cite{Luscher:1992an,Luscher:1993gh} and the twisted boundary conditions~\cite{deDivitiis:1993hj,Itou:2012qn}) on hypertorus (${\mathbb{T}^4}$).
Otherwise, the classical solution does not connect to the standard perturbative vacuum~\cite{tHooft:1981nnx, tHooft:1979rtg,Luscher:1985wf} because of a gauge inequivalent configuration, which is called {\it toron}, of the degenerate minimal action~\cite{GonzalezArroyo:1981vw,Coste:1985mn}.

On the other hand, as a proper spacetime structure, a large aspect ratio between the spatial and temporal directions might be a good setup to consider a well-defined theory in the weak coupling regime, according to the recent studies of the resurgence scenario.
It is well-known that the perturbative expansion of the SU($N_c$) gauge theory on ${\mathbb{R}^4}$ spacetime does not converge in higher order calculations.
The resurgence scenario has been proposed~\cite{Bogomolny:1980ur, ZinnJustin:1981dx} to solve this problem for the quantum mechanical models and low-dimensional quantum field theories, which suffer from a similar problem.
In the scenario, {\it a compact dimension and/or a boundary condition with ${\mathbb{Z}}_N$-holonomy} are introduced as a proper choice of the spacetime structure.
On the modified spacetime, the perturbative series and the nonperturbative effects of the theories are related to each other in the weak coupling regime, and physical quantities are determined without any imaginary ambiguities.
A characteristic phenomenon in the nonperturbative side of this scenario is the appearance of local topological objects with a {\it fractional} charge, which contributes to the perturbative vacua, as a (semi)-classical solution of the theories.
The fractionality of the topological charges is needed to cancel the renormalon pole~\cite{'tHooft:1977am,Beneke:1998ui} whose action is of order $1/N_c$ in comparison with the action for the integer-instantons~\cite{Argyres:2012ka}. 
Recently, the resurgence structure has been revealed in several quantum-mechanical models and low-dimensional quantum field theories~\cite{Argyres:2012ka,Dunne:2012ae,Dunne:2012zk,Dunne:2013ada,Misumi:2015dua,Dunne:2016jsr,Misumi:2014jua,Fujimori:2016ljw,Fujimori:2017oab,Fujimori:2017osz}.
 A signal of the fractionality of the energy density has been observed in the Principle Chiral Model using the lattice numerical simulation~\cite{Buividovich:2017jea}.

For the SU($N_c$) gauge theory, a recent paper~\cite{Yamazaki:2017ulc} has proposed a promising regularization formula on $\mathbb{T}^3 \times \mathbb{R}$.
The authors pointed out that the IR cutoff is necessary, which should be higher energy than the dynamical IR scale, namely $\Lambda$ scale in the SU($N_c$) gauge theory, otherwise, the trans-series expansion of physical observables breaks down.
Therefore, they introduce the twisted boundary conditions for the {\it{two}} compactified dimensions using the center symmetry.
The twisted boundary conditions induce the IR cutoff in the gluon propagator, and the fractional-instanton is allowed as a  nonperturbative object even in the weak coupling regime~\cite{tHooft:1981nnx, tHooft:1979rtg, deDivitiis:1993hj,Witten:1982df}.
Furthermore, the center symmetry can be dynamically restored because of the tunneling behavior between $Z_{N_c}$-degenerate vacua.
It seems to be promising to discuss the adiabatic continuity, where no phase transition occurs toward the decompactified limit in contrast to the first order phase transition at the finite-temperature.
Although the resurgence structure of the SU($N_c$) gauge theory on the modified spacetime has not yet been proven, these phenomena are very similar to the ones in low-dimensional models, which are successfully resurgent.

\begin{figure}[h]
\centering\includegraphics[width=13.5cm]{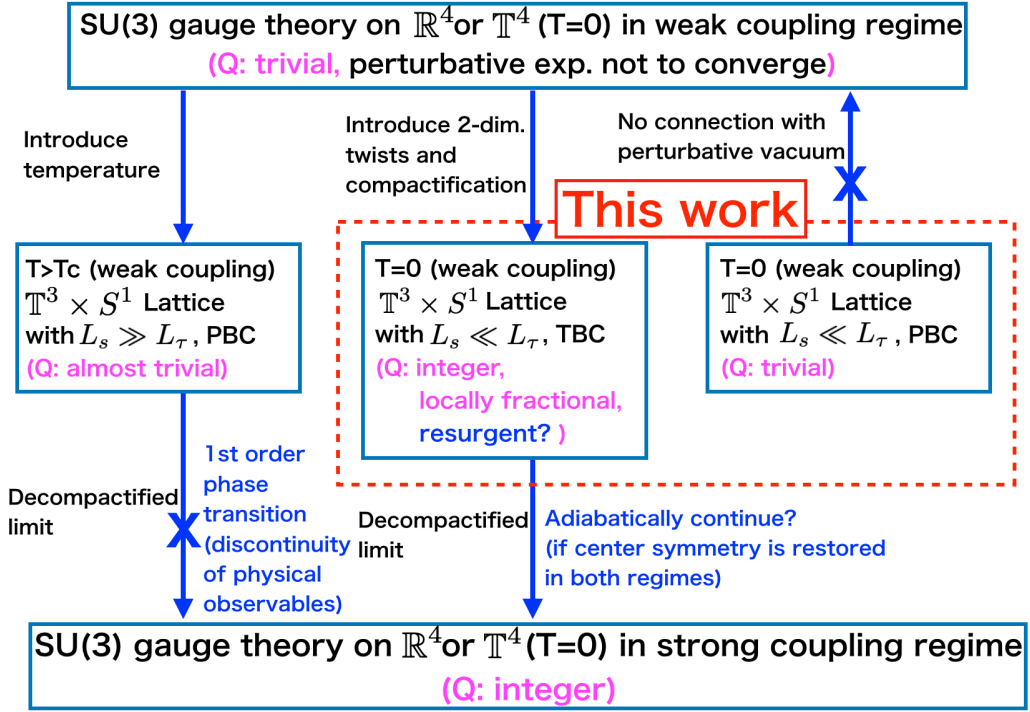}
\caption{Relationships among the SU($3$) gauge theories on various spacetime structures and coupling regimes. The topological property ($Q$) for each setup is also shown. 
Cross symbols on the arrow denote that two theories does not connect with each other smoothly. 
In this work, we focus on two boxes located at the center and the center-right.}
\label{fig:Image}
\end{figure}
Based on these situations, 
we perform the lattice numerical simulation on  $\mathbb{T}^3 \times S^1$ with the twisted boundary conditions to study the topology in the weak coupling regime.
Here, we maintain a large aspect ratio between two radii for the three-dimensional torus ($L_s$) and the temporal circle ($L_\tau$) as $L_s \ll L_\tau$ (center box in Fig.~\ref{fig:Image}).
The boundary conditions and the spacetime structure used here are equivalent to those in Refs.~\cite{Yamazaki:2017ulc,Witten:1982df} in the continuum and $S^1 \to {\mathbb R}$ limits.
We tune lattice parameters to be in a weak coupling regime, where the running coupling constant under the same twisted boundary conditions has already studied in Ref.~\cite{Itou:2012qn}.
For a comparison purpose, we also perform the simulations on the ordinary periodic lattice with the same lattice parameters (center-right box in Fig.~\ref{fig:Image}).
To study the fractional topology and its basic properties of the SU($3$)  gauge theory is interesting itself.
Furthermore, it is a first step for the challenge to define a regularized gauge theory from the perturbative to the nonperturbative regimes without any obstructions.

To confirm whether the configurations with fractional charges are generated, we directly measure the topological charge of configurations on the twisted lattice.
We find that  the configurations with nonzero $Q$ appear on the twisted lattice even in the weak coupling regime.
Although the total topological charge, $Q$, takes an integer value as in the strong coupling regime on the ordinary periodic lattice, some of them consist of multiple fractional-instantons.
We also study the relationship between the fractional-instantons and other nonperturbative phenomena, namely the tunneling, the center symmetry, and the confinement.
We show that the fractional-instanton connects two of the degenerate $\mathbb{Z}_{N_c}$-broken vacua, that is the same properties with the fractional-instantons discussed in Refs.~\cite{Yamazaki:2017ulc,Witten:1982df}.
Investigating the scaling law of the Polyakov loop, we find that although the center symmetry seems to be partially restored, the free-energy of single probe quark is still finite, which  is consistent with the deconfinement.

This paper is organized as follows:
In \S.~\ref{sec:twisted-bc}, we give a review of the basic properties of the lattice gauge theory with the twisted boundary conditions for the two compactified directions.
We give comments on the absence of the zero-modes and the existence of the fractional-instantons as a classical solution.
In \S.~\ref{sec:strategy}, the strategy of the Monte Carlo simulation is explained. We tune the simulation parameters to be in the perturbative regime ($g^2 \approx 0.7$) and take a sufficiently large extent to generate multi-instanton configurations.
We also describe our sampling method for a tiny lattice spacing with a long autocorrelation.
Section~\ref{sec:results} presents the simulation results. In \S.~\ref{sec:topology}, the configurations with nonzero $Q$ show up on the twisted lattice in the weak coupling regime, while there is no such configuration on the ordinary periodic lattice with the same lattice parameters.
We find the local topological object with a fractional charge in the configurations with nonzero $Q$.
The distribution of the Polyakov loop on the twisted lattice has the different behavior from that on the periodic lattice as shown in \S.~\ref{sec:Ploop}.
We also show the tunneling phenomena between the $\mathbb{Z}_{N_c}$-degenerate vacua by studying the local topological charge and the Polyakov loop on each site in \S.~\ref{sec:relation}.
In \S.~\ref{sec:deconfined}, the deconfinement property of these configurations is discussed.
The last section contains the summary and several future directions.

\section{Twisted boundary conditions on hypertorus lattice }\label{sec:twisted-bc}
\subsection{Twisted boundary conditions and  the absence of zero-modes}
We review the properties of the SU($N_c$) gauge theory on the lattice with the two-dimensional twisted boundary conditions.
On the hypertorus with the ordinary periodic boundary conditions, the saddle points of the action are not only degenerate through gauge transformations, but an extra degeneracy exists due to the global toroidal structure~\cite{GonzalezArroyo:1981vw,Coste:1985mn}. 
The configuration is related to a part of the zero-modes and is called toron.
On the other hand, the two-dimensional twisted boundary conditions eliminate all zero-modes, so that it is not suffered from the toron problem.
Here we explain the detail setup and show the gluon propagator is regularized by an IR momentum cutoff because of the twisted boundary conditions~\cite{Luscher:1985wf, deDivitiis:1993hj, Itou:2012qn}.

Let us start with the Wilson-Plaquette action for the SU($N_c$) gauge theory; 
\beq
S_W=  \frac{2{N_c}}{g_0^2} \sum_{n, \mu > \nu}  \left(1- \frac{1}{{N_c}} \mbox{Tr} P_{\mu \nu} (n) \right).\label{eq:Wilson-action}
\eeq
Here, $g_0$ and $P_{\mu \nu}$ denote the lattice bare coupling constant and the plaquette,
\beq
P_{\mu \nu}(n)=U_{\mu} (n) U_{\nu} (n+\hat{\mu}) U^\dag_{\mu}(n+\hat{\nu}) U^\dag_\nu (n),
\eeq
respectively.
$U_\mu(n)$ represents a link variable from a site $n=(n_x,n_y,n_z,n_\tau)$ to its neighbor in the $\mu$-direction, and takes values with the SU($N_c$) group elements.

We introduce the twisted boundary conditions in the $x$ and $y$ directions;
\beq
U_{\mu}(n+\hat{\nu} N_s)&=&\Omega_{\nu} U_{\mu}(n) \Omega^{\dag}_{\nu} \mbox{ for $\nu=x,y$}, \label{twisted-bc-gauge} \\
\Omega_\nu &\in& \mbox{SU(}N_c\mbox{)},\nonumber
\eeq
while the ordinary periodic boundary conditions are imposed in the $z$ and $\tau$ directions;
\beq
U_{\mu}(n+\hat{\nu} N_s)= U_{\mu}(n)  \mbox{ for $\nu=z,\tau$}.  \label{eq:untwisted-bc}
\eeq
Here, $N_s$ denotes the lattice extent for each direction in lattice unit and is related with the size of the torus ($L_s = a N_s$).
For simplicity, only in this subsection, we consider the lattice extents for all directions have the same length.
$\Omega_{\nu}$ ($\nu=x,y$) are the twist and SU($N_c$) matrices.
In the case of $N_c=3$, they have the following properties,
\beq
 \Omega_{\nu} \Omega_{\nu}^{\dag}=\mathbb{I},
 (\Omega_{\nu})^3=\mathbb{I},
 \mbox{Tr}[\Omega_{\nu}]=0, ~
\mbox{and} ~~
\Omega_{\mu}\Omega_{\nu}=e^{i2\pi/3}\Omega_{\nu}\Omega_{\mu},
\eeq
for a given $\mu$ and $\nu$($\ne \mu$).

At a corner on the lattice in the $x$-$y$ plane, a translation of a link variable is given by
\beq
U_\mu (n +\hat{x}N_s + \hat{y}N_s)= \Omega_x \Omega_y U_\mu (n) \Omega_y^\dag \Omega_x^\dag.
\eeq
The interchanging $\Omega_x$ and $\Omega_y$ in this equation leads to the same result.
The gauge transformation for the original link variable;
\beq
U_\mu (n) \rightarrow \Lambda(n) U_\mu (n) \Lambda^\dag (n+ \hat{\mu})
\eeq
implies 
\beq
\Lambda (n+\hat{\nu} N_s) = \Omega_\nu \Lambda (n) \Omega^\dag_\nu .\label{eq:org-gauge-trans}
\eeq
Then, the Wilson-Plaquette action with the twisted boundary conditions at the boundary, for instance $n_y=N_s-1$, is given by
\beq
P_{xy} &=& U_x(n_x,N_s-1,n_z,n_\tau) U_y (n_x+1,N_s-1,n_z,n_\tau) \Omega_y U^\dag_x (n_x,0,n_z,n_\tau) \nonumber\\
&& \times \Omega^\dag_y U^\dag_y (n_x,N_s-1,n_z,n_\tau).
\eeq
The toron configurations, which are related to the closed winding around the whole torus, are not transformed into themselves by the twisted conditions~\cite{GonzalezArroyo:1981vw}. 
They do not have a degenerate energy with the standard vacuum, since the plaquette on the boundary gives a different contribution from the standard one.

Next, we consider the gluon propagator and show that it has an IR cutoff in this lattice setup.
The link variable can be parameterized by the gauge fields ($A_\mu(n)$) as
\beq
U_\mu (n) = e^{i g_0 A_\mu (n)},~~~\mbox{with }
A^\dag_\mu (n) = A_\mu (n),~~~~ \mbox{Tr}[A_\mu (n)] =0.
\eeq
The corresponding boundary conditions for gauge field $A_\mu (n)$ imply 
\beq
A_{\mu} (n+ \hat{\nu} N_s)&=& \Omega_\nu A_\mu (n) \Omega^\dag_\nu, \quad \nu = x,y, \label{eq:twc-A-xy}\\
A_{\mu} (n+ \hat{\nu} N_s)&= &A_\mu (n) , \quad \quad ~  \quad \nu = z,\tau.\label{eq:twc-A-zt}
\eeq
The plane-wave expansion of the gauge field is given by
\beq
A_\mu (n) = \frac{1}{N_s^4} \sum_k \Gamma_k \tilde{A}_\mu (k) e^{ikn +i k_\mu/2},\label{eq:plane-wave}
\eeq
where $\Gamma_k$ is a $N_c \times N_c$ complex matrix.
Substituting Eq.~(\ref{eq:plane-wave}) to Eq.~(\ref{eq:twc-A-xy}) gives
\beq
  e^{ik_\nu N_s} \Gamma_k= \Omega_\nu \Gamma_k \Omega^\dag_\nu,~\label{eq:Gamma-k}
\eeq
for $\nu = x,y$.
The non-zero solution is realized only if the momentum components satisfy
\beq
k_{x,y} &=& k_{x,y}^{ph}+k_{x,y}^{\perp},~~~~ k_{z,\tau}=k_{z,\tau}^{ph},
\eeq
where
\beq
k_\mu &=&\frac{2\pi m_{\mu}^{ph}}{N_s}, ~~~~~k_\mu^\perp = \frac{2 \pi m_{\mu}^{\perp}}{N_c N_s}.
\eeq
Here, $-N_s/2 \le m_\mu^{ph} < N_s/2$ denotes the degree of freedom for the ordinary physical momentum for $\mu=x,y,z,$ and $\tau$, while there is an additional unphysical degree of freedom $m_{\mu}^{\perp}=0,1,N_c-1$ for the twisted directions.
There is a one-to-one correspondence between the unphysical momenta and the color degrees of freedom of $A_\mu$~\cite{deDivitiis:1993hj}.
Actually, the number of the combination of $(m_x^\perp, m_y^\perp)$ is ${N_c}^2-1$ due to the traceless condition of the gauge field.

We can carry out perturbative calculations using those unphysical momenta instead of the color degrees of freedom.
If we take the Feynman gauge, then the gluon propagator in the momentum space is given by
\beq
\langle \tilde{A}_\mu (q^{ph},q^\perp) \tilde{A}_\nu (k^{ph},k^\perp) \rangle = \frac{1}{2{N_c}} \delta^{(4)}_{(q+k)^{ph},0} \delta^{(2)} _{(q+k)^{\perp},0} (1- \delta^{(2)}_{k^\perp,0}) e^{- \pi i (k^\perp,k^\perp)/3 } \frac{1}{k^2} \delta_{\mu \nu},\nonumber\\
\eeq
where
$k^2=4 \sum_{\mu} \sin^2 (k_\mu/2)$ and $(\tilde{k}^\perp,k^\perp)=\tilde{m}^\perp_x m^\perp_x + \tilde{m}^\perp_y m^\perp_y + (\tilde{m}^\perp_x + \tilde{m}^\perp_y)(m^\perp_x + m^\perp_y)$.
Because of the factor  $(1- \delta^{(2)}_{k^\perp,0})$, the zero-modes including the torons are excluded in the propagator.
It corresponds to introducing the IR cutoff proportional to $1/(N_c N_s)$ in momentum space.

\subsection{Classical solutions with twisted boundary conditions}\label{sec:classical-solution}
If the four-dimensional twisted boundary conditions are introduced on the hypertorus ($\mathbb{T}^4$), it is known that the topological charge can be a fraction~\cite{tHooft:1979rtg, tHooft:1981nnx};
\beq
Q= \frac{1}{32 \pi^2} \int d^4 x \mbox{Tr} \epsilon_{\mu \nu \rho \sigma} F_{\mu \nu} F_{\rho \sigma} = \mbox{integer} - \frac{\kappa}{N_c}. \label{eq-tHooft-Q}
\eeq
Here, $\kappa=\frac{1}{8}\epsilon_{\mu \nu \rho \sigma} n_{\mu \nu} n_{\rho \sigma}$, and $n_{\mu \nu}$  has six integers ($n_{\mu \nu} = - n_{\nu \mu}$) defined modulo $N_c$ and labels the twist;
\beq
\Omega_\mu \Omega_\nu = e^{2 \pi i n_{\mu \nu}/N_c} \Omega_\nu \Omega_\mu.
\eeq
The boundary conditions correspond to the introduction of the magnetic flux for all four-directions. To see the fractional-instantons, the strength of the magnetic flux (or size of the torus) for each direction has to be tuned (see Eq.~(4.19) in Ref.~\cite{tHooft:1981nnx}). 
It shows that $Q$ is always integer if we introduce the twisted boundary conditions only in two dimensions, since $n_{12}= - n_{21}=1$ and the others are zero.

On the other hand, it is known that even though only two dimensions have the twisted boundary conditions, the classical solution with the fractional topological charge emerges on $\mathbb{T}^3 \times \mathbb{R}$ (see also section $7$ in~\cite{Witten:1982df}). 
There are $N_c$ degenerate classical vacua, and a fractional-instanton appears as a configuration connecting between them~\cite{Yamazaki:2017ulc}.
In our simulations, if we take the infinite size limit of the temporal direction, then the spacetime structure is the same with the one in Refs.~\cite{Yamazaki:2017ulc,Witten:1982df}.
Then, we expect that the similar fractional-instantons as these works could emerge as a classical solution in the numerical simulations.
Therefore, it is worth to summarize the origin of the fractional topological charge and the properties of the fractional-instantons on $\mathbb{T}^3 \times \mathbb{R}$.

When we consider the perturbative expansion around the classical solution, we have to take not only $U_\mu= \mathbbm{1}$ but also its gauge equivalent configurations.
For simplicity, we take pure gauge and consider $A_\mu  = - \Lambda (\partial_\mu \Lambda)$.
The gauge field $A_\mu$ satisfies the twisted boundary conditions Eqs.~(\ref{eq:twc-A-xy}) and~(\ref{eq:twc-A-zt}).
In previous subsection, we consider the naive boundary condition, Eq.~(\ref{eq:org-gauge-trans}), for $\Lambda$, but the boundary conditions for $A_\mu$ are still satisfied even if the following extended $\mathbb{Z}_{N_c}$ gauge transformations are added;
\beq
\Lambda(n+ \hat{x}N_s) &=& e^{2 \pi i l_x/N_c} \Omega_x \Lambda (n) \Omega_x^{\dag}, \nonumber\\
\Lambda(n+ \hat{y}N_s) &=& e^{2 \pi i l_y/N_c} \Omega_y \Lambda (n) \Omega_y^{\dag}, \nonumber\\
\Lambda(n+ \hat{z}N_s) &=& e^{2 \pi i l_z/N_c}  \Lambda (n) , \label{eq:extended-TBC}
\eeq
where $l_x,l_y,$ and $l_z$ are integers (modulo $N_c$).
We denote these $\mathbb{Z}_{N_c}$ gauge transformations with $(l_x,l_y,l_z)= (1,0,0),(0,1,0),$ and $(0,0,1)$ as $T_x,T_y,$ and $T_z$, respectively.
Then, any gauge transformation $\Lambda$ with the boundary conditions, Eq.~(\ref{eq:extended-TBC}), can be written by
\beq
\Lambda = (T_x)^{l_x} (T_y)^{l_y} (T_z)^{l_z} \tilde{\Lambda}, \label{eq:extended-gauge-trans}
\eeq
where $\tilde{\Lambda}$ satisfy the original gauge transformation Eq.~(\ref{eq:org-gauge-trans}).
$T_x$ and $T_y$ can be chosen as a constant matrix, and then $(T_x)^{N_c}=(T_y)^{N_c}=1$.
However, $(T_z)^{N_c}$ can not be continuously deformed to the identity in all spacetime coordinates, and shifts the topological charge.

To see that, we put the topological charge (Eq.~(\ref{eq:def-Q}))  in
\beq
Q &=& \frac{1}{8 \pi^2}  \int \mbox{Tr} ( F \wedge F ), \nonumber\\
&=& - \frac{1}{24 \pi^2} \int \mbox{Tr} (\Lambda^{-1} d \Lambda)\wedge (\Lambda^{-1} d \Lambda) \wedge (\Lambda^{-1} d \Lambda). \label{eq:Q-wedge}
\eeq
Utilizing $(\Lambda^{-1} d \Lambda) = d (\ln \Lambda )$ and substituting Eq.~(\ref{eq:extended-gauge-trans}) to Eq.~(\ref{eq:Q-wedge}) imply
\beq
Q&=&
 \frac{l'}{N_c} + \mbox{integer} .
\eeq
Here, $l' = l_z \cdot n'$, where $n'$ appears since the logarithmic function is a multi-valued function.
Thus, if $l_z$ and/or $n'$ are not a multiple of $N_c$, then a fractional topological charge is allowed on $\mathbb{T}^3 \times \mathbb{R}$.

Furthermore, the extended $\mathbb{Z}_{N_c}$ gauge transformation could rotate the Polyakov loop in the $z$-direction in the complex plane.
Let us consider the Polyakov loop in the $z$-direction;
\beq
P_z = \frac{1}{N_c} \mbox{Tr} \exp \left[ i \int A_z dz \right].
\eeq
Then, the Polyakov loop for the gauge equivalent, $A_z \rightarrow \Lambda^{-1} A_z \Lambda -i \Lambda^{-1} (\partial_z \Lambda)$, is given by
\beq
\frac{1}{N_c} \mbox{Tr} \exp \left[ i \int (\Lambda^{-1} A_z \Lambda - \Lambda^{-1} (\partial_z \Lambda) )dz \right]&=&\frac{1}{N_c} \mbox{Tr} \exp \left[ i \left( \int A_z  dz + 2\pi l_z /N_c+ 2\pi n \right) \right],  \nonumber\\
 &=& e^{2\pi i l_z /N_c} P_z.
\eeq
If $l_z$ is not a multiple of $N_c$, then the phase factor remains in $P_z$.
It is therefore possible that if the classical solution with fractional topological charge appears at some spacetime coordinate, then the complex phase of  $P_z$ rotates at the same coordinate.

The other typical property of the fractional-instantons on $\mathbb{T}^3 \times \mathbb{R}$ is an absence of the size-modulus~\cite{Yamazaki:2017ulc}.
The IR cutoff is also related to the existence of the fractional instanton on $\mathbb{T}^3 \times \mathbb{R}$.
According to Ref.~\cite{Yamazaki:2017ulc}, all size-moduli of the integer-instanton on $\mathbb{R}^4$ are associated with the translation moduli of the fractional-instanton on $\mathbb{T}^3 \times \mathbb{R}$.
Then, the fractional-instanton with the smallest topological charge ($Q=1/{N_c}$) has no size modulus, and hence the size of fractional-instantons is unique.
The size is related to the compactification radius ($L_s$) of $\mathbb{T}^3$.
The intuitive understanding of the absence of the size-modulus is the following\footnote{We appreciate K.~Yonekura for useful discussions.};
if the size of instanton is smaller than the compactification radius, then the instanton becomes the ordinary integer-instanton, since the situation is the same as in the $\mathbb{R}^4$ spacetime.
On the other hand, the fractional-instantons with a large size is also forbidden, since $A_\mu$ has a non-zero ``mass'' coming from the unphysical momenta ($k^{\perp}\propto 1/L_s$) of the twisted directions.
Then, the size of the fractional-instanton with the smallest charge is fixed.
The fractional-instanton with larger charges can be constructed from the composite states of the smallest one.

In this work, we consider $\mathbb{T}^3 \times S^1$, where the radius of $\mathbb{T}^3$ is smaller than the one of $S^1$ ($L_\tau$),
and introduce the two-dimensional twisted boundary conditions into $\mathbb{T}^3$.
The total topological charge must be integer, since $\kappa=0$ in Eq.~(\ref{eq-tHooft-Q}) and the homotopy class of $\mathbb{T}^3 \times S^1$ is the same with the one of $\mathbb{T}^4$.
However, if the instanton size is much smaller than $L_\tau$, the spacetime structure, that the instanton feels, can be approximated by $\mathbb{T}^3 \times \mathbb{R}$.
Then, the emergent of several fractional-instantons is locally allowed as a classical solution, if the sum of these topological charges is an integer.
We may see the evidence of the fractional-instantons from the rotation of the Polyakov loop and the uniqueness of the size of the fractional-instantons in numerical simulations.

\section{Simulation strategy}\label{sec:strategy}
\subsection{Lattice parameters}
The simulation strategy to investigate the nonperturbative properties in the perturbative regime is as follows.
We utilize the Wilson-Plaquette gauge action given by Eq.~(\ref{eq:Wilson-action}) as a lattice gauge action.
We put the lattice parameter $\beta=2{N_c}/g_0^2$ with $N_c=3$.
The other lattice parameters, which we can control by hand, are the lattice extents in spatial ($N_s$) and temporal directions ($N_\tau$). The lattice spacing ``$a$'' is put unity during the numerical simulation. 
Once we introduce the physical quantity as a reference scale, for instance, the Sommer scale~\cite{Sommer:1993ce} and $t_0$ scale in the gradient flow method~\cite{Luscher:2010iy}, then we obtain a one-to-one correspondence between $\beta$ and ``$a$'', since the SU($3$) gauge theory has  only one dynamical scale.

To investigate a nontrivial semi-classical solution, we set the lattice parameters ($\beta, N_s, N_\tau$) to satisfy the following three conditions;\\
(1) the twisted boundary conditions on the two spatial dimensions to introduce the IR cutoff and to kill the torons \\
(2) sufficiently large lattice extent to generate multiple topological objects  \\
(3) tuned lattice gauge coupling to realize the perturbative regime 

To satisfy the condition (1), we use the $N_s^3 \times N_\tau$ lattice, where the twisted boundary conditions is imposed on the $x$ and $y$ directions in the three spatial directions. 
The $z$-direction has the same lattice extent with the $x,y$-directions, but its boundary condition is periodic.

The size of the fractional-instanton is predicted to be the same with the compactification radius ($N_s$). 
Then, at least one-direction (here $N_\tau$) is larger than $N_c\times N_s$, to satisfy the condition (2). 
We choose $(N_s,N_\tau) =(12, 60)$ for this work.

For the condition (3), we take $\beta=16$.
According to Fig.~$4$ in Ref.~\cite{Itou:2012qn}, the running coupling constant at the scale ($L_s= a N_s$) indicates $g^2(1/L_s) \approx 0.7$.
It is consistent well with the result of the $1$--loop approximation, which is independent of the renormalization scheme.
If we fix the $\Lambda$ scale where the running coupling constant in the Twisted-Polyakov-Loop scheme diverges as shown in Ref.~\cite{Itou:2012qn}, the lattice setup with ($\beta,N_s$) $=$ ($16, 12$) corresponds to $L_s \Lambda \approx 1.5^{-24}$.
It satisfies the Dunne-\"{U}nsal  condition, ${N_c} L_s \Lambda \ll 2\pi$, where it is expected that the system is in the weak coupling regime but still there are some nonperturbative features.
The lattice spacing is $a \approx 5.0 \times 10^{-6}$ [fm], if we use $\Lambda=200$ [MeV].
Although the size of this lattice is extremely small, the small lattice would be suitable to study the semi-classical behavior in the weak coupling regime.
Actually, the action density ($S_W/N_s^3N_\tau$) is roughly $0.048$ 
in ($\beta, N_s,N_\tau$)=($16,12,60$), which is close to the classical limit, where the action takes a minimum value.

\subsection{Sampling method of the configurations in high $\beta$}
To collect the gauge configurations in a weak coupling regime, we have to take care of the autocorrelation; a newly generated configuration, which is updated from the old configuration using the random numbers, is very similar to the old one. 
The autocorrelation length depends on observables, and generally, quantities related to the low-modes have a long autocorrelation.
 The autocorrelation length of the topological charge is a few ten- or hundred-sweeps (see {\it e.g.}~\cite{Schaefer:2010hu}) in a typical calculation for the SU($3$) gauge theory at the zero-temperature with $a \approx O(10^{-2})$[fm].
Since the length grows in proportion to $O(1/a^5)$~\cite{Schaefer:2010hu,DelDebbio:2002xa,Luscher:2010we}, the simulations with $a \approx 5.0 \times 10^{-5}$ must suffer from a severe autocorrelation problem.

To avoid this, we prepare the $100$ seeds of random-number generation, here we label them as $\#1$ -- $\#100$.
We independently update $O(10^3)$ sweeps using each random-number series.
Here, we call one sweep as a combination of one Pseudo-Heat-Bath (PHB) update and $10$ over-relaxation steps.
We collect $100$ configurations as samples in a fixed $N$-th sweep and label the samples ``conf.$\#$'' using its seed of the random-number series.
For the comparison, we also generate the other $100$ configurations using the same method and the same lattice parameters, while the boundary conditions are periodic for four directions.
From now on, we use the term ``TBC lattice'' to denote the lattice with ($x,y,z,\tau$) $=$ (twisted, twisted, periodic, periodic) boundaries, while the term ``PBC lattice''  denotes the one with the periodic boundaries for all directions.

In the end of this section, let us show the explicit form of the twist matrices in our numerical calculations for $N_c=3$~\cite{Itou:2012qn,Trottier:2001vj},
\beq
\Omega_x=\left( 
\begin{array}{ccc}
0 & 1 & 0\\
0 & 0 & 1\\
1 & 0 & 0
\end{array}
\right),
\Omega_y=\left( 
\begin{array}{ccc}
e^{-i2\pi /3} & 0 & 0\\
0 & e^{i2\pi /3} & 0\\
0 & 0 & 1
\end{array}
\right).\label{eq:twist-mat}
\eeq 

\section{Results}\label{sec:results}
\subsection{Topological charge}\label{sec:topology}
We measure the topological charge, Eq.~(\ref{eq:def-Q}), by using the clover-leaf operator on the lattice (see Eq.(2.18) in Ref.~\cite{Luscher:1996sc}).
\begin{figure}[h]
\vspace{0.5cm}
\centering\includegraphics[width=10cm]{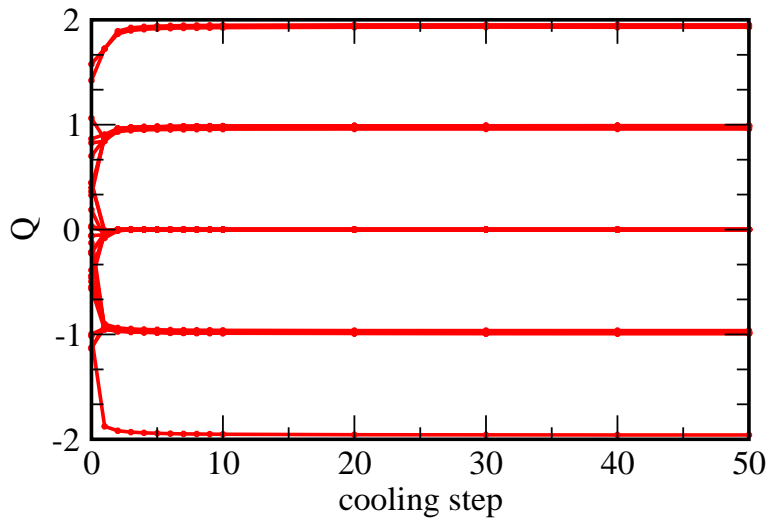}
\caption{Cooling step dependence of the topological charge on the TBC lattice.}
\label{fig-total-Q-icool}
\end{figure}
The topological charge of the gauge configuration generated by Monte Carlo simulations usually does not take an integer value because of UV fluctuations.
We utilize the cooling method~\cite{Berg:1981nw, Iwasaki:1983bv,Teper:1985rb}, which is an evolving step to minimize the gauge action by smoothing the configurations.
Figure~\ref{fig-total-Q-icool} presents $Q$ as a function of the cooling steps for several configurations (conf.\# $1$ -- $30$) in the TBC lattice.
$Q$ rapidly converges to an (almost) integer value with a few cooling steps.
We perform the cooling until $200$ steps and confirm that the plateau continues.
Here, the discrepancy from an integer value, at most $(\Delta Q/Q) \approx 0.04$, comes from lattice artifacts.
In this paper, we neglect the small discrepancy and approximate the value of $Q$ in the plateau of the cooling steps to an integer value.
Now, we fix the number of cooling steps as $50$ ($N$-cool $=50$) and the number of sweeps as $2000$.

\begin{figure}[h]
\vspace{0.5cm}
\centering\includegraphics[width=13cm]{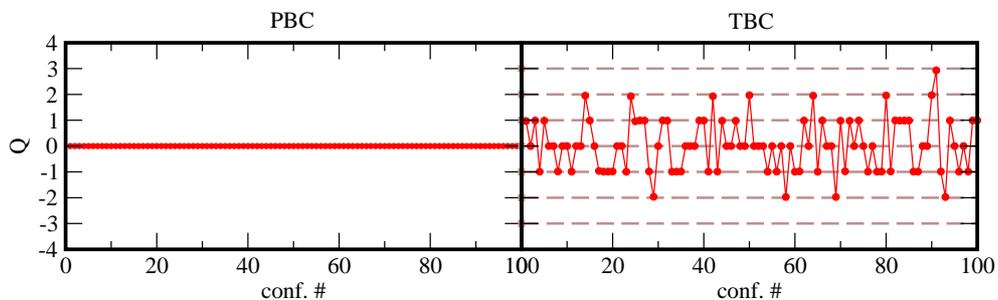}
\caption{Total topological charge ($Q$) on the PBC (left) and the TBC (right) lattices for each of $100$ configurations.}
\label{fig-total-Q-PBC-TBC}
\end{figure}

The left (right) panel of Fig.~\ref{fig-total-Q-PBC-TBC} shows the result on the PBC (TBC) lattice.
The horizontal axis denotes the configuration-number labeled by the seed of random-series.
All configurations have $Q=0$ on the PBC lattice,
while some configurations have non-zero $Q$ on the TBC lattice. 
$Q$ is distributed over $-2 \le Q \le 3$, and the number of configurations with non-zero $Q$ is $66$ while the remaining $34$ configurations live in the $Q=0$ sector on the TBC lattice.

\begin{figure}[h]
\centering\includegraphics[width=10cm]{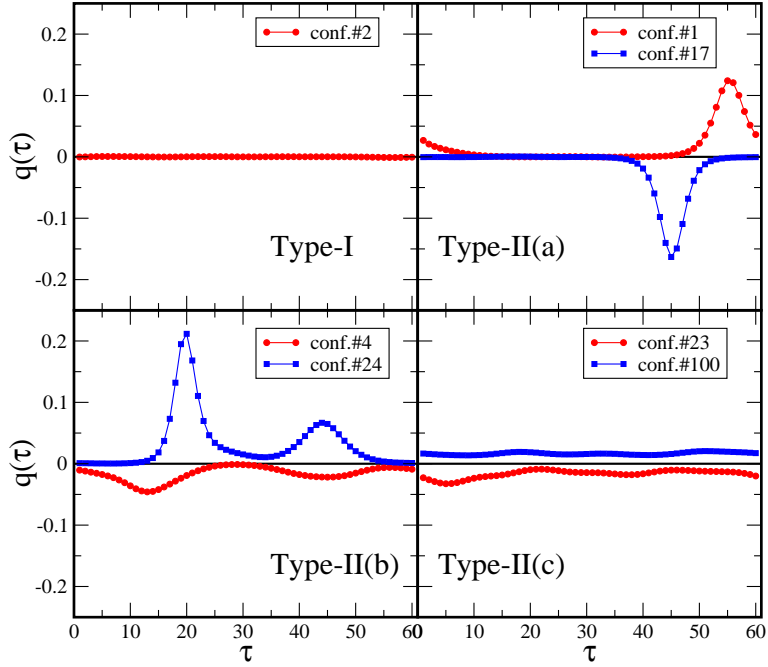}
\caption{Typical distributions of the local topological charges ($q(\tau)$). The integer-instanton and integer-anti-instanton are shown in the top-right panel. On the other hand, the topological charges are ``fractionalized''  in the bottom-left panel (see Eq.~(\ref{eq:ex-frac-q})).}
\label{fig-local-q-tau-TBC}
\end{figure}
Now, let us focus on the results on the TBC lattice.
We classify the configurations into two types: {\bf{Type-I}} for $Q=0$ and {\bf{Type-II}} for $Q \neq 0 $.
Furthermore, we find that the magnitude of topological charge on each lattice site strongly depends on $\tau$ in several configurations.
Taking the sum for the three-dimensional spaces, we define a local charge,
\beq
q(\tau) = \frac{1}{32 \pi^2} \sum_{x,y,z} \mbox{Tr} \epsilon_{\mu \nu \rho \sigma}  F_{\mu \nu} F_{\rho \sigma} (x,y,z,\tau).
\eeq
We plot the local charge for several typical configurations in Fig.~\ref{fig-local-q-tau-TBC}.
The top-left panel shows the local charge of the configuration in {\bf Type-I} (conf.\#2).
We find that $q(\tau)$ is always zero for any $\tau$ for all configurations in this type. 
{\bf Type-II} configurations are further classified into three types as follows;\\
\begin{description}
\item[\bf Type-II(a)] it has a single peak (the top-right panel)
\item[\bf Type-II(b)]  it has several peaks (the bottom-left panel)
\item[\bf Type-II(c)]  it takes a continuous non-zero value (the bottom-right panel)
\end{description}
In the case of {\bf Type-II(a)}, the sum of $q(\tau)$ around the single peak agrees with the value of $Q$. 
For instance, the confs.$\# 1$ (red-circle) and $\#17$ (blue-square) have $Q=+1$ and $Q=-1$, respectively.
These peaks can be interpreted as the integer-instanton and integer-anti-instanton, respectively.
In the case of {\bf Type-II(c)}, we cannot see an excess of $q(\tau)$ in spite of the fact that the sum of $q(\tau)$ for all $\tau$ gives a nonzero integer.
For instance, the sum of $q(\tau)$ in the conf.$\# 23$ (read-circle) is $Q=-1$, and the one in conf. $\# 100$ (blue-square) is $Q=+1$.
We find a uniform behavior for the $z$-direction of the local charge on site-by-site, which is similar to that for the action density in the Principal Chiral Model given in Ref.~\cite{Buividovich:2017jea}.
The local charge of  {\bf Type-II(c)} configurations is of the same order on all sites in contrast to the appearance of more than $O(10^2)$ difference in {\bf Type II(a)} and {\bf Type II(b)}. 
This suggests that there is no local excess of $q(\tau)$ in the whole spacetime coordinate in  {\bf Type-II(c)} configurations.

The configurations in {\bf Type-II(b)} are the most interesting.
We can take the sum of $q(\tau)$ around each peak by dividing whole the domain of $\tau$ into several domains, whose boundaries are defined by the local minimum of $\vert q(\tau) \vert$.
Then, each sum is proportional to $n/3$ within $\Delta Q/Q \approx 0.04$ error, where $n$ is an integer except for a multiple of $3$.
The confs.$\# 4$ (red-circle) and $\# 24$ (blue-square) plotted in Fig.~\ref{fig-local-q-tau-TBC} have the total instanton number $Q=-1$ and $Q=+2$, respectively.
We find
\beq
\mbox{conf.\#4 } ~~~Q_1&=& \sum_{\tau=29}^{55} q(\tau)=-0.343 \approx -\frac{1}{3}, \nonumber\\
~Q_2 &=& Q-Q_1=  -0.647 \approx  -\frac{2}{3},\nonumber\\
\mbox{conf.\#24 } 
~Q_1 & =& \sum_{\tau=6}^{33} q(\tau)= 1.269  \approx  \frac{4}{3},\nonumber\\
~Q_2 &=& Q-Q_1=0.656 \approx \frac{2}{3}.\label{eq:ex-frac-q}
\eeq
Thus, some integer-instantons contain multiple fractional-instantons in the weak coupling regime.
This is the first obeservation of the fractional-instantons of the SU($3$) gauge theory on the deform spacetime using the lattice simulation.

\begin{figure}[!h]
\vspace{0.1cm}
\centering\includegraphics[width=13cm]{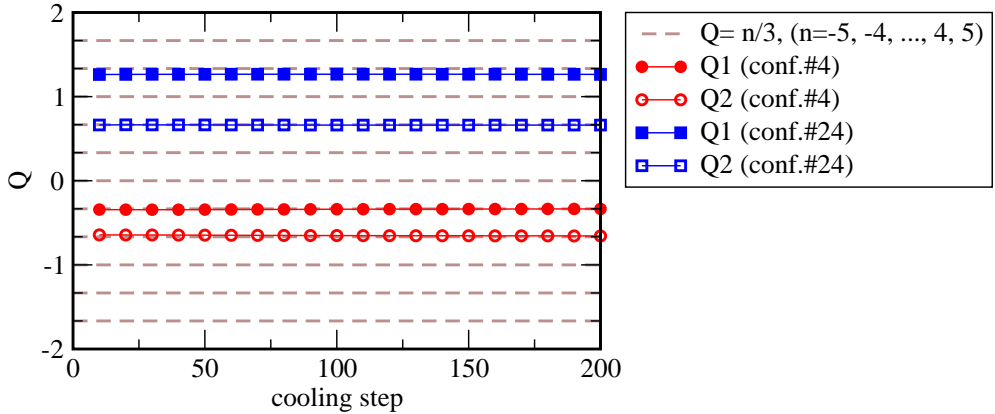}
\caption{Cooling step dependence of the local charges, $Q_1$ (filled symbol) and $Q_2$ (open symbol), for  confs.~$\# 4$ (red-circle) and $\# 24$ (blue-square).}
\label{fig-icool-deps}
\end{figure}
To confirm that the fractionality of the charge is not a quantum fluctuation, 
let us investigate the stability of local fractional-instantons during the cooling processes.
If it is just a quantum fluctuation then the charge would disappear by the cooling process.
Figure~\ref{fig-icool-deps} displays the local charges ($Q_1,Q_2$) as a function of the cooling steps for confs.~$\#4$ (red-circle) and $\#24$ (blue-square).
We find that the position of the local minimum of $\vert q(\tau) \vert$ is independent of the cooling steps.
Then, each charge ($Q_1$,$Q_2$) is very stable.

Next, we investigate the topology changing during the Monte Carlo updates.
In ordinary lattice calculations for the SU($3$) gauge theory in the strong coupling regime, the total topological charge can change within a few ten or hundred Monte Carlo sweeps, since the potential barrier is finite on the lattice.
In general, the topology changing does not so frequently occur if the lattice spacing is very tiny on the PBC lattice.
However, we find that a changing of the local topological charge rather frequently occurs on the TBC lattice, and then $Q$ can also change its value during the Monte Carlo updates.

\begin{figure}[!h]
\centering\includegraphics[width=15cm]{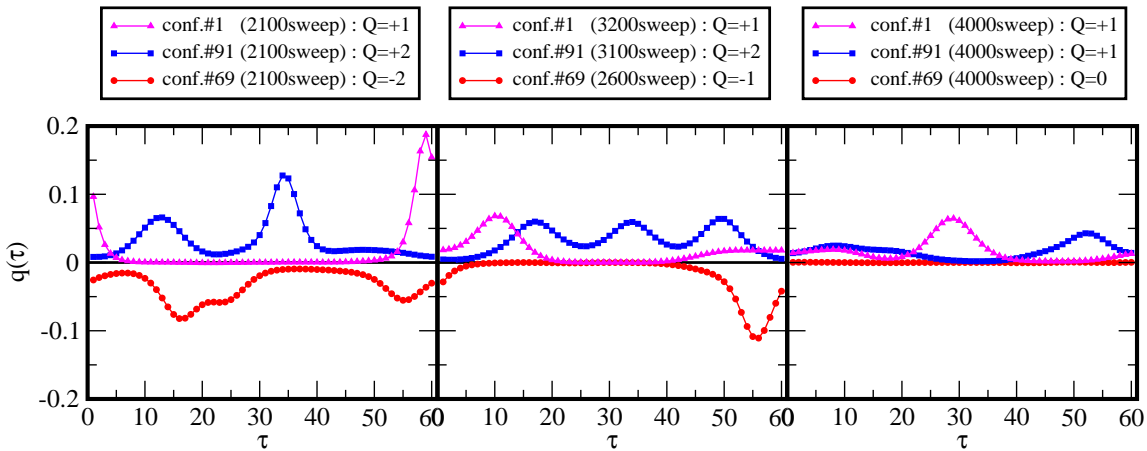}
\caption{Example of the sweep dependence of the local topological charge ($q(\tau)$). 
The magenta-triangle, blue-square, and red-circle symbols denote the confs.~$\#1, \# 91$, and $\#69$, respectively.   }
\label{fig-local-q-conf69}
\end{figure}
Typical results for the topology changing are shown in Fig.~\ref{fig-local-q-conf69}.
In all panels, the number of cooling processes is fixed to $50$.
During the Monte Carlo updates from the $2100$-th to $4000$-th sweep, the total charge changes as follows;
\beq
\mbox{conf.}\#1 && Q\mbox{ does not change ($Q=+1$)} , \nonumber\\
\mbox{conf.}\#91 && Q=+2 \rightarrow Q=+2 \rightarrow Q=+1, \nonumber\\
\mbox{conf.}\#69 && Q=-2 \rightarrow Q=-1 \rightarrow Q=0. \nonumber
\eeq
Meanwhile, the combination of the local charges shows rich variations;
\beq
\mbox{conf.}\#1 && (+1 \mbox{ with single peak}) \rightarrow (+2/3,+1/3) \rightarrow (+2/3,+1/3), \nonumber\\
\mbox{conf.}\#91 && (+2/3,+4/3) \rightarrow (+2/3,+2/3,+2/3) \rightarrow (+1/3,+2/3), \nonumber\\
\mbox{conf.}\#69 && (-4/3,-2/3) \rightarrow (-1 \mbox{ with single peak}) \rightarrow  (q(\tau)=0). \nonumber
\eeq
Thus, multiple fractional-instantons merge into an integer-instanton and vice versa, and a fractional-instanton with a large charge deforms into multiple fractional-instantons with a small charge. 

It is known by the analytical study on the $\mathbb{C}P^{N-1}$ model in low dimensions that the fractional-instantons can transform into the integer-instanton and vice verse if the moduli-parameter is changed by hand (see Fig.~$4$ in Ref.~\cite{Eto:2006mz}). 
If two fractional-instantons approach to each other and merge into one, then the translation moduli of the fractional-instanton are back to the size-moduli of the integer-instanton.
On the other hand, if the size of an integer-instanton is larger than the radius of the compactified direction, then the integer-instanton is divided into several fractional-instantons.
We can conclude that a similar phenomenon dynamically occurs during the Monte Carlo simulations of the SU($3$) gauge theory.

We put three remarks;
the first one is related to a bion configuration.
Among $2100$ configurations we investigated, there is no $Q=0$ configuration containing a pair of the fractional-instanton and the fractional-anti-instanton.
Such a configuration is called bion, which plays an important role to see the resurgence structure in the $\mathbb{C}P^{N-1}$ model~\cite{Fujimori:2016ljw,Fujimori:2017oab}.
We consider that the absence of bions would come from the interaction between the fractional-instanton and fractional-anti-instanton and the finite volume effect.

The second one is a distribution of the topological charge in the weak coupling regime.
In our calculation, we find a decrease in the topological charge in the long Monte Carlo sweeps.
The numbers of configurations in the $Q=0$ and $Q \ne 0$ sectors are $34$ and $66$ at $2000$-th sweep, respectively,
and become $49$ and $51$ at $4000$-th sweeps.
We cannot find the configuration whose $\vert Q \vert$ increases during these Monte Carlo sweep.
It might suggest that only the $Q=0$ sector is preferred after the infinitely long updates that would be related to the probability weight of the topological sector.
On the other hand, we believe that the fractional-instantons will be stabilized in the $L_\tau \rightarrow \infty$ limit.
The determination of the topological susceptibility in $\mathbb{T}^3 \times \mathbb{R}$ must be interesting and will be carried out by the simulations with a large volume and a large aspect ratio ($L_\tau/L_s$).

The third one is for the size-modulus of the fractional-instantons.
According to Ref.~\cite{Yamazaki:2017ulc}, there is no size-modulus of the fractional-instantons.
Ten fractional-instantons with $Q=\pm 2/3$ are plotted in Figs.~\ref{fig-local-q-tau-TBC} and \ref{fig-local-q-conf69}.
It seems that they all have almost the same shape, namely a similar height of the peak and a similar curve around it.
To be precise, the peak-height takes $|q(\tau)|=0.04 \sim 0.07$.
We consider that they are consistent with each other because there is a $\Delta Q/Q \approx 0.04$ error in our simulations.
On the other hand, some fractional-instantons with the other charges have a broader width.
The loss of the uniqueness of the size for the fractional-instantons would come from the periodicity of the temporal direction ($S^1$) in our simulation compared with $\mathbb{R}$.
It deforms the shape of the fractional-instantons to satisfy the constraint, that the sum of them takes an integer value.
The simulations with a larger extent for the temporal direction and the continuum extrapolation could improve the situation, and reveal the uniqueness of the shape for the fractional-instantons.

\subsection{Polyakov loop and center symmetry}\label{sec:Ploop}
The Polyakov loop ($P$) is an order parameter of the center symmetry breaking, and behaves as $P \rightarrow Pe^{2\pi i k/3}$ with $k=0,1,$ and $2$, under the center transformation. 
On the PBC lattice, the Polyakov loop in the $\mu$-direction is given by the product of the link variables of the direction;
\beq
P_{\mu} = \frac{1}{V} \sum_{\mbox{sites of }\nu (\ne \mu)} \frac{1}{N_c} {\mbox{Tr}} \left[ \prod_{j} U_{\mu} (\mu=j,\nu) \right]. \label{eq:def-pbc-Poly}
\eeq
Here, we take the spacetime average for each configuration, and $V=N_s^3$ for $\mu=\tau$ and $V=N_s^2  N_\tau$ for the others.
On the other hand, the definition of the Polyakov loop in the twisted directions on the TBC lattice are modified as,
\beq
P_{x} =  \frac{1}{N_s^2 N_\tau} \sum_{y,z,\tau}  \frac{1}{N_c} {\mbox{Tr}} \left[ \prod_{j} U_{x}(x=j,y,z,\tau) \Omega_{x} e^{i2\pi y/3L} \right],\label{eq:def-twisted-Poly}
\eeq
in order to satisfy the gauge invariance and the translational invariance.

\begin{figure}[h]
\vspace{0.3cm}
\centering\includegraphics[width=10cm]{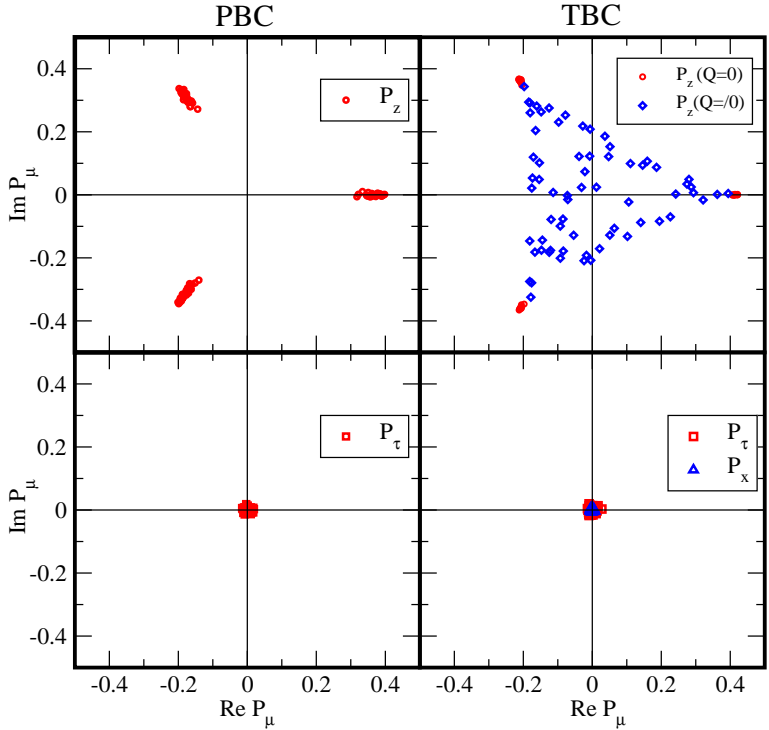}
\caption{Scatter plot of the Polyakov loop for each direction. Top-left and bottom-left panels show the spatial and the temporal Polyakov loops on the PBC lattice, respectively.  Top-right panel shows the results of $P_z$ , while the bottom-right panel gives the ones of $P_x$ (blue-triangle) and $P_\tau$ (red-square) on the TBC lattice. }
\label{fig-scatter}
\end{figure}
The scatter plots of the Polyakov loop in each direction are given in Fig.~\ref{fig-scatter}.
Each point denotes the data for one configuration.
Here, all configurations are at $2000$-th sweep from the random start.  
The results for the PBC lattice are shown in the left panels.
Since, the $x,y$, and $z$ directions are equivalent, we present the Polyakov loop only in the $z$ and $\tau$  directions in the top-left and  bottom-left panels, respectively.
At $\beta=16$ with $N_s=12$, it is clearly in the deconfinement phase because of its scale ($L_s\Lambda \approx 6.0 \times 10^{-5} $).
Then, the Polyakov loop in the $z$-direction is located at one of three degenerate vacua, whose complex phases are $0$ and $\pm 2\pi/3$.
In the continuum limit with fixed physical lattice-size, one of three vacua is chosen, and therefore the center symmetry is spontaneously broken, which is the same as in the situation of the SU($3$) gauge theory in the high-temperature.
The Polyakov loop in the $\tau$ direction seems to be invariant under the center transformation since they are located around the origin.

On the other hand, the right panels in Fig.~\ref{fig-scatter} show the Polyakov loop on the TBC lattice.
The Polyakov loop in the $x$-direction, where the twisted boundary condition is imposed, is shown in the right-bottom panel.
Because of the twist matrix, the Polyakov loop is located around the origin in the complex plane.
The result for the $y$-direction is the same as the one for the $x$-direction.
For the $\tau$-direction, the behavior is the same as the one on the PBC lattice.
$P_z$ on the TBC lattice shows a curious behavior, even though the boundary condition for the $z$-direction is periodic.
$P_z$ is spread over almost the whole triangle, where the product of the link variables before taking the trace in Eq.~(\ref{eq:def-pbc-Poly}) satisfies the unitarity condition.
The location of each data is changed under the center transformation if $|P_z| \ne 0$, so that the center symmetry in most configurations is broken in the same meaning as in the case of the finite-temperature.
However, we find that its breaking is milder than the one on the PBC lattice because the average of $\vert P_z \vert $ is smaller.

Note that in the top-right panel of Fig.~\ref{fig-scatter}, the red-circle symbols located in one of the $\mathbb{Z}_3$-degenerate vacua denote the configurations with $Q=0$, while the blue-diamond symbols inside the triangle correspond to the configurations with $Q \ne 0$.
The figure clearly suggests that there is a relationship between the value of $Q$ and the Polyakov loop in the $z$-direction.

\subsection{Tunneling phenomena and fractional instanton}\label{sec:relation}
Now, let us investigate the relationship shown at the end of the previous subsection.
We introduce the Polyakov loop in the $z$-direction on each lattice site;
\beq
\tilde{P}_{z} (x,y,\tau) &=& \frac{1}{N_c} {\mbox{Tr}} \left[ \prod_{j} U_{z} (x,y,z=j,\tau) \right], \nonumber\\
&\equiv& |\tilde{P}_z(x,y,\tau)| e^{i\varphi (x,y,\tau)}.\label{eq:def-phi}
\eeq
\begin{figure}[h]
\centering\includegraphics[width=10cm]{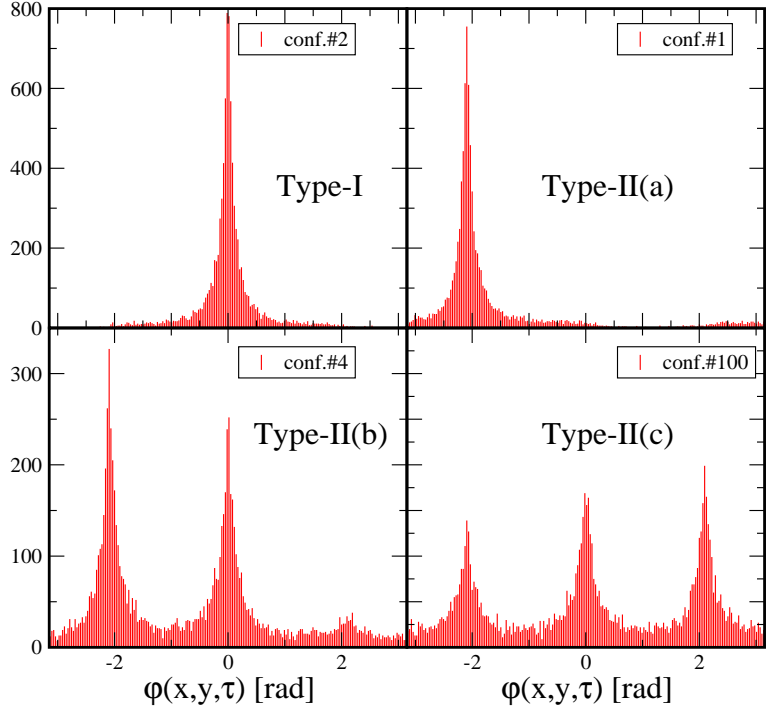}
\caption{Histograms of $\varphi(x,y,\tau)$ for typical configurations, which are classified by the local charge ($q(\tau)$). The corresponding data of the local charge are shown in Fig~\ref{fig-local-q-tau-TBC}.}
\label{fig:hist}
\end{figure}
The histograms of $\varphi(x,y,\tau)$ for typical configurations are shown in Fig.~\ref{fig:hist}.
Here, the corresponding data of the local charge are displayed in Fig.~\ref{fig-local-q-tau-TBC}.
In the case of {\bf Type-I} and {\bf Type-II(a)}, $\tilde{P}_z(x,y,\tau)$ on all sites are located at one of the ${\mathbb Z}_3$-degenerate vacua.

\begin{figure}[!h]
\centering\includegraphics[width=10cm]{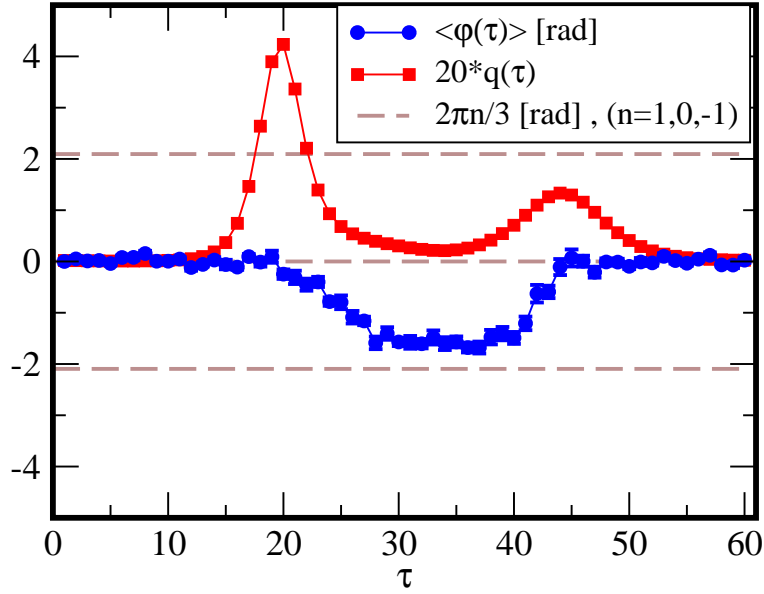}
\caption{$\tau$-dependence of the averaged complex phase (blue-circle) and the local topological charge (red-square) for conf.$\# 24$. }
\label{fig-Q-Ploopz}
\end{figure}

On the other hand, in the case of {\bf Type-II(b)} configurations, two of the ${\mathbb{Z}_3}$-degenerate vacua are chosen.
To see the manifest relationship between the fractional-instanton and the distribution of the Polyakov loop, we plot the averaged complex phase $\langle \varphi (\tau) \rangle$, which is defined by $\langle \varphi (\tau) \rangle \equiv \sum_{x,y} \varphi(x,y,\tau)/N_s^2$, for conf.$\# 24$ as a function of $\tau$ (the blue-circle symbols) in Fig.~\ref{fig-Q-Ploopz}.
We also present the local topological charge $q(\tau)$ as the red-square symbols, where it is multiplied by $20$ so as to be easily seen.
We find that $\langle \varphi \rangle$ starts changing its value around the peak of the local charge ($q(\tau)$), where the fractional-instanton exists.
This indicates that the fractional-instanton is related to the rotation of the complex phase of $P_z$.
That is the same as the properties of the classical solutions on $\mathbb{T}^3 \times \mathbb{R}$ as shown in \S.~\ref{sec:classical-solution}.
We can therefore conclude that the fractional-instantons on $\mathbb{T}^3 \times \mathbb{R}$ are obtained by the numerical simulations on $\mathbb{T}^3 \times S^1$.

Furthermore, it shows the relation between the tunneling among the $\mathbb{Z}_3$-degenerate vacua and the fractional-instantons.
It has been also discussed in the context of the two-dimensional $\mathbb{C}P^{N-1}$ model.
The model is obtained from the dimensional reduction of the four-dimensional SU($N$) gauge theory with the twisted boundary conditions~\cite{Yamazaki:2017ulc,Yamazaki:2017dra,Tanizaki:2017qhf, Eto:2006mz,Eto:2004rz,Bruckmann:2007zh,Brendel:2009mp}.
In the limit where the ($x,z$) directions shrink, the four-dimensional SU($N$) gauge theory is reduced to the two-dimensional nonlinear sigma model, whose boundary condition in the compactified direction ($y$) has the $\mathbb{Z}_{N}$-holonomy. 
Non-zero expectation values of the Polyakov loop for the shrinking direction ($z$) correspond to the vacuum expectation value (v.e.v.) of the complex scalar field in the reduced theory, where the v.e.v. depends on $\tau$.
The fractional-instanton can be interpreted as a classical solution connecting two vacua with different v.e.v. of the complex scalar field.
Our numerical results in Fig.~\ref{fig-Q-Ploopz} show that similar phenomenon occurs in the fractional-instantons of the four-dimensional SU($N$) gauge theory.

In the case of {\bf Type-II(c)}, the histogram of $\varphi (x,y,\tau)$ has three peaks at three degenerate vacua equally.
Here, we find that no clear $\tau$-dependence exists in its distribution.
We expect that the tunneling phenomena among three vacua occur also through $x$ and $y$ directions. 
Because the magnitude of the Polyakov loop given in Eq.~(\ref{eq:def-pbc-Poly}) is very small ($|P_z| \ll 0.1$), the Polyakov loops ($P_\mu$) in all directions are located near the origin in the complex plane.
That means the center symmetry is dynamically restored.
Such a dynamical restoration of the center symmetry is predicted in Ref.~\cite{Yamazaki:2017ulc} on ${\mathbb{T}}^3 \times {\mathbb{R}}$ spacetime.
In our numerical calculation on ${\mathbb{T}}^3 \times S^1$ lattice, the configuration {\bf Type-II(c)}  is rare: three per one-hundred.
If {\bf Type-II(c)} is dominant in the continuum and/or the $S^1 \rightarrow {\mathbb{R}}$ limits, then the center symmetry would be completely preserved even in weak coupling regime. 
It is an important future work to find which type of configurations remains in these limits.

\subsection{Polyakov loop and confinement}\label{sec:deconfined}
In this section, we focus on the other nonperturbative phenomenon: the confinement.
We have seen from the Polyakov loops in the $x,y,$ and $\tau$ directions that the center symmetry seems to be restored.
The center-symmetric distribution of the Polyakov loops in the $x$ and $y$ directions comes from the twist matrices in Eq.~(\ref{eq:def-twisted-Poly}).
On the other hand, the Polyakov loop in the $\tau$-direction also indicates the center-symmetric property, though it is possible to show the spontaneous symmetry-breaking. 
Generally, the Polyakov loop is related to the free-energy of single (probe) quark, $\langle |P_\tau| \rangle \propto e^{-N_\tau F_q}$.
In the confinement phase, $F_q$ is large and diverges in the infinite-volume limit, so that  $\langle |P_\tau| \rangle \sim 0$ can be naively interpreted as a confinement. 
However, we need to  find whether the smallness of $\vert P_\tau \vert$ comes from a large $F_q$ or a large $N_\tau$ with a finite value of $F_q$, since we take a large lattice extent ($N_\tau=60$) with an extremely small lattice spacing.
Here, we will confirm the deconfinement property of our configurations on the TBC lattice even though these configurations exhibit $|P_\tau| \sim 0$.

\begin{figure}[h]
\vspace{0.5cm}
\centering\includegraphics[width=10cm]{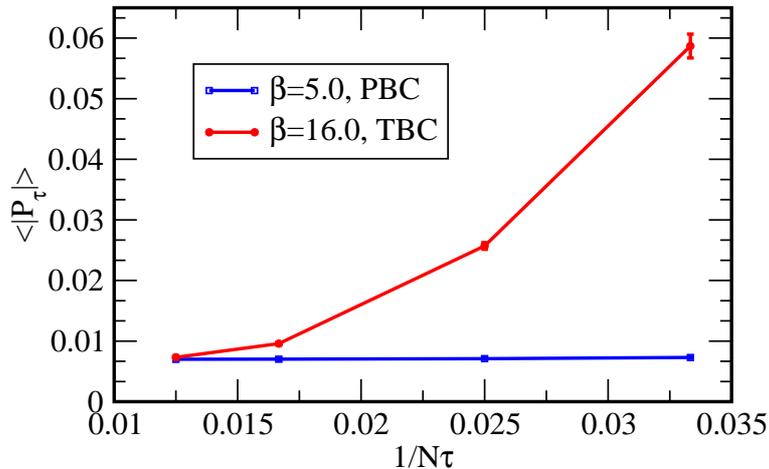}
\caption{Magnitude of Polyakov loop in the $\tau$-direction as a function of $N_\tau$. We take $N_\tau=30,40,60$, and $80$ from right to left.}
\label{fig:ABS-Ploopt}
\end{figure}
Figure~\ref{fig:ABS-Ploopt} shows $\vert P_\tau \vert$ as a function of $1/N_\tau$ for two lattice parameters: $(\beta,N_s)=(16, 12)$ on the TBC lattice and $(\beta,N_s)=(5,12)$ on the PBC lattice.
It is known that the latter case exhibits the confinement since the critical temperature is determined as $\beta_c=6.3384$ in $(N_s,N_\tau)=(\infty, 12)$ in the finite-temperature simulations~\cite{Boyd:1996bx}. 
The blue-square symbols denote the results for the PBC lattices.
All data in $30 \le N_\tau \le 80$ agree with each other within $2$-$\sigma$ statistical error bar, which implies no $N_\tau$-dependence.
This is a natural consequence in the confinement phase.
On the other hand, $\vert P_\tau \vert$ on the TBC lattice (red-circle symbols) decreases as increasing $N_\tau$~\footnote{We cannot directly compare the absolute values of Polyakov loop between $\beta=5$ and $\beta=16$,  since the lattice raw data have to be multiplicatively renormalized, where the renormalization factor depends on the value of $\beta$~\cite{Gupta:2007ax}. }. 
The data at $N_\tau=60$, which we mainly focus on in this work, is still in the middle of its decreasing beyond the statistical error bar.
We conclude that the configurations with the fractional-instanton have the deconfinement property in the present lattice setup.

\section{Summary and future works}\label{sec:conclusion}
We have studied the nonperturbative phenomena of the SU($3$) gauge theory in the weak coupling regime on $\mathbb{T}^3 \times S^1$ with the large aspect ratio between two radii. 
This is the first work to find a fractional-instanton in the weak coupling regime and its nonperturbative properties in the Monte Carlo simulations on a promising deformed spacetime toward the resurgence of the SU($3$) gauge theory.
Introducing the twisted boundary conditions into two directions realizes the perturbative standard vacuum on the hypertorus and is related to the existence of the fractional-instantons. 
We can conclude that the fractional-instantons in this work  have the same properties as the ones of the classical solutions given by the gauge equivalent of the standard perturbative vacua under the extended $\mathbb{Z}_3$ gauge symmetry in the $S^1 \rightarrow \mathbb{R}$ limit.

The numerical results show that the total topological charge ($Q$) always takes integer-values including nonzero on the TBC lattice, while it is fixed to only zero on the PBC lattice.
On the TBC lattice, there are four types of configurations depending on its distribution of the local topological charge ($q(\tau)$); it takes zero in all $\tau$ ({\bf Type-I}), it exhibits a integer-instanton ({\bf Type-II(a)}), it includes multiple fractional-instantons ({\bf Type-II(b)}), and it is continuously nonzero ({\bf Type-II(c)}).
We find that the fractional-instantons can merge into the integer-instanton and vice versa during the Monte Carlo update processes.
On the other hand, they are stable during a cooling process, and therefore we conclude that the fractionality is not a quantum fluctuation. 

We have also investigated the center symmetry by observing the Polyakov loop for each spacetime direction.
The Polyakov loop in the $z$-direction on the TBC lattice shows the different behavior from the one on the PBC lattice, though the same boundary condition is applied to the direction. 
The Polyakov loop is scattered over the unitary triangle in the complex plane.
Configurations are located at one of the $\mathbb{Z}_3$-degenerate vacua for $Q=0$, while the Polyakov loops live inside the unitary triangle for $Q\ne 0$.
We have shown that the averaged complex phase of $P_z$ rotates if the fractional-instanton emerges.
Thus, fractional-instantons connect two of the $\mathbb{Z}_3$-degenerate phases of the Polyakov loop in the $z$-direction.
It is the same property as the one of the classical solution on $\mathbb{T}^3 \times \mathbb{R}$.
On the other hand, the Polyakov loop in the $\tau$-direction seems to be center-symmetric, but its scaling property indicates the deconfinement property.
Furthermore, we have found that the configurations of {\bf Type-II(c)} exist, whose Polyakov loops in all directions exhibit the center-symmetric property even in the weak coupling regime.

According to the analogy of the quantum mechanical models and the low-dimensional quantum field theories,  the existence of the fractional-instantons will give an additional contribution to physical observables in the weak coupling regime and will solve the imaginary-ambiguity problem of the perturbative expansion.
Furthermore, the center-symmetric property even in the weak coupling regime is promising to show the adiabatic continuity between the weak and strong coupling regimes.
We believe that these phenomena in the weak coupling regime, which are found in this work, will play an important role to study the resurgence structure of the SU($3$) gauge theory.

We address the future works and related lattice works as follows:\\
\indent {\bf Resurgence structure of the SU($3$) gauge theory}\\
To see a resurgence structure, we have to investigate at least three points in future: (i) finding a fractional topological object which gives a contribution to an physical observable ({\it {e.g.}} plaquette) in the $Q=0$ sector (ii) comparing the contribution between  the renormalon pole~\cite{Bali:2014fea} in the perturbative series and the nonperturbative background with the fractional topology (iii) seeing the adiabatic continuity~\cite{Sulejmanpasic:2016llc, Bergner:2018unx} to the decompactified limit.
It is also an important work to see how to take the decompactified limit with keeping the resurgence structure and contributions coming from the fractional-instantons.

{\bf Including the dynamical quarks: $\mathbb{Z}_{N_c}$-QCD and adjoint QCD}\\
We expect that a similar topological object with the fractional charges appears in the QCD-like theories including the dynamical fermions.
It is known that there are at least two promising models: the $\mathbb{Z}_{N_c}$-QCD~\cite{Kouno:2015sja,Iritani:2015ara} and the adjoint QCD models~\cite{Bergner:2018unx,Cossu:2009sq,Cossu:2013ora,Cossu:2013nla}.
The formulation of the two-dimensional twists of this work can be extended to systems including dynamical fermions~\cite{Parisi:1984cy}.
The advantage of the usage of the twisted boundary conditions is not only the absence of toron but also the induced IR momentum cutoff.
Hence, we can perform simulations with exact massless fermions.
It must be helpful to investigate the adiabatic continuity near the massless limit as discussed in Ref.~\cite{Bergner:2018unx}.

{\bf Other lattice calculations to find a fractional instanton: \\
Schr\"{o}dinger functional boundary, the four-dimensional twisted, and other approaches}\\
It might be worth to mentioning the other formulations that show the fractional-instantons on the lattice.

A similar discussion to this work might be possible by using the lattice setup with the Schr\"{o}dinger functional boundary~\cite{Luscher:1992an} with a large aspect ratio between the spatial and temporal extents.
To kill all zero-modes and to stabilize fractional-instantons may also need an additional gauge fixing or the other technique~\cite{Aoki:1998qd}.

We can also consider the twisted boundary conditions for three or four directions.
Although it has been discussed in the strong coupling regime, the lattice numerical simulations with the four-dimensional twists have successfully been carried for the SU(${N_c}$) gauge theories~\cite{GarciaPerez:1989gt, deForcrand:2002vs}.
Note that the theory with four-dimensional twisted boundary conditions locally has the same gauge symmetry with SU(${N_c}$), but the global symmetry becomes SU(${N_c}$)$/\mathbb{Z}_{N_c}$.
Furthermore, the fractional topologies at the finite- and zero-temperature with a nontrivial holonomy have been investigated~\cite{Lee:1998vu,Lee:1998bb, Kraan:1998kp,Kraan:1998pm,Gattringer:2002tg,Bruckmann:2003yq, Bruckmann:2004ib,Ilgenfritz:2005um}.
The energy density and the zero-mode density of the calorons for the SU($2$) and SU($3$) gauge theories have been numerically shown. 
The other approaches to find the fractional topological charges on the lattice have been done by using the Dirac operator with the higher-dimensional representations on the periodic lattices~\cite{ Edwards:1998dj,Fodor:2009nh,Kitano:2017jng}.


\acknowledgments

The author appreciates S.~Aoki and K.~Yonekura for deep discussions about the lattice numerical data, and would like to thank T.~Fujimori, T.~Misumi, M.~Nitta, N.~Sakai, Y.~Tanizaki and M.~Yamazaki for valuable comments. 
We are also indebted to S.~Aoki, T.~Misumi, K.~Nagata and N.~Sakai for helpful comments on this manuscript.
The author is grateful to the organizers and participants of “RIMS-iTHEMS International Work-shop on Resurgence Theory” at RIKEN, Kobe for giving her a chance to deepen their ideas.  
In particular, the talk given by K.~Yonekura at the workshop and the private conversation were invaluable to start this work.
Numerical simulations were performed on SX-ACE at the Research Center for Nuclear Physics (RCNP), Osaka University. 
This work is supported by the Ministry of Education, Culture, Sports, Science, and Technology (MEXT)-Supported Program for the Strategic Research Foundation at Private Universities “Topological Science” (Grant No. S1511006),
and is also supported in part by the Japan Society for the Promotion of Science (JSPS) Grant-in-Aid for Scientific Research (KAKENHI) Grant Number 18H01217. 



\end{document}